\documentclass[prd, preprint, nofootinbib, showpacs]{revtex4}

\usepackage{graphicx}
\usepackage{amsfonts}
\usepackage{latexsym}
\usepackage{amssymb}
\usepackage{epsfig}

\begin{document}

\title{
Gravitino dark matter from increased thermal relic particles 
}

\author{Nobuchika Okada}
 \email{okadan@post.kek.jp}
 \affiliation{
Theory Division, KEK, 1-1 Oho, Tsukuba 305-0801, Japan 
}

\author{Osamu Seto}
 \email{osamu.seto@uam.es}
 \affiliation{
 Instituto de F\'{i}sica Te\'{o}rica UAM/CSIC,
 C-XVI, Universidad Aut\'{o}noma de Madrid,
 Cantoblanco, Madrid 28049, Spain
}

%

\begin{abstract}

We investigate the so-called superWIMP scenario 
 with gravitino as the lightest supersymmetric particle (LSP) 
 in the context of non-standard cosmology, in particular, 
 brane world cosmology. 
As a candidate of the next-to-LSP (NLSP), 
 we examine slepton and sneutrino. 
Brane world cosmological effects dramatically enhance 
 the relic density of the slepton or sneutrino NLSP, 
 so that the NLSP with mass of order $100$ GeV  
 can provide the correct abundance of gravitino dark matter 
 through its decay. 
We find that with an appropriate five dimensional Planck mass, 
 this scenario can be realized consistently 
 with the constraints from Big Bang Nucleosynthesis (BBN) 
 for both NLSP candidates of slepton and sneutrino. 
The BBN constraints for slepton NLSP are more stringent 
 than that for sneutrino, as the result, 
 the gravitino must be rather warm in the slepton NLSP case. 
The energy density of gravitino produced by thermal scattering 
 is highly suppressed and negligible 
 due to the brane world cosmological effects.

\end{abstract}


\preprint{KEK-TH-1189} 
\preprint{IFT-UAM/CSIC-07-43} 

\vspace*{3cm}
\maketitle


\section{Introduction}

Various cosmological observations have revealed 
 the ingredients and their abundance in the Universe, 
 and showed that 
 the standard $\Lambda$CDM cosmological model agrees with these data well.
In fact, the abundance of cold dark matter is also precisely measured as
\begin{equation}
\Omega_{CDM} h^2 \simeq 0.11 ,
\label{OmegaCDM}
\end{equation}
 by the Wilkinson Microwave Anisotropy Probe (WMAP) satellite~\cite{WMAP}.
However, the identity and the origin of dark matter particles
 are still open questions in cosmology and particle physics.

The lightest supersymmetric particle (LSP) in supersymmetric models 
 with the conserved R-parity is suitable for cold dark matter.
Well-studied candidates are the lightest neutralino, gravitino, and axino.
Historically the lightest sneutrino also has been considered 
 as a dark matter candidate~\cite{Ibanez:1983kw}, but 
 it was excluded because of its relic abundance being too small 
 and, conclusively, the null result 
 in the direct dark matter searches~\cite{Falk:1994es}
 \footnote{Sneutrino here means a left-handed (active) one 
 in the minimal supersymmetric standard model (MSSM) with
 the lepton number conservation. The case with the lepton number violation
 is discussed in Ref.~\cite{HMM}.
 A right-handed (sterile) sneutrino~\cite{RHsneutrino}
 or a mixture of left- and right-handed sneutrino~\cite{Mixedsneutrino}
 is still viable.}.

Presumably, among them, the lightest neutralino 
 is the most attractive candidate from various points of view. 
In more general terms, the weakly interacting massive particle (WIMP) 
 is the attractive dark matter candidate, 
 which the neutralino belongs to. 
The most interesting feature of the WIMP dark matter 
 is that it can be thermal relics 
 and its abundance is described as~\cite{Kolb} 
\begin{equation}
\Omega_{DM}h^2 \simeq 1 \times 10^9
 \frac{(m_{DM}/T_f) \rm{GeV}^{-1}}
 {\sqrt{g_*} M_{Pl} \langle\sigma v\rangle} ,
 \label{StandardThAbundance}
\end{equation}
 where $m_{DM}$ is the mass of the dark matter particle,
 $T_f$ is the freeze out temperature, 
 $g_*$ is the effective total number of relativistic degrees of freedom,
 $M_{Pl} = 1.2 \times 10^{19}$ GeV is the Planck mass, and
 $\langle\sigma v\rangle$ is the thermal averaged product of
 the annihilation cross section and the relative velocity.
This relic abundance is essentially determined 
 by only its annihilation cross section, 
 and does not depend on the detail thermal history 
 of the early Universe before $T \gtrsim 10$ GeV. 
This predictability and definiteness is fascinating,
 compared with the most of models for nonthermal production 
 of dark matter where the resultant dark matter relic density 
 is highly model dependent. 
In addition, on-going or forthcoming experiments 
 may observe such a dark matter directly and/or indirectly.

However, there is another interesting possibility 
 for the candidate of dark matter 
 with highly suppressed interactions. 
Even though it cannot be in thermal equilibrium, 
 this type of dark matter can have the appealing feature 
 in its abundance. 
This is called the superWIMP scenario, 
 where the superWIMP dark matter is produced through 
 the decay of a long-lived WIMP~\cite{Covi:1999ty, Feng:2003xh} 
 with highly suppressed interactions between 
 the WIMP and the superWIMP. 
In this case, the dark matter abundance is given as 
\begin{equation}
\Omega_{DM}h^2 = \frac{m_{DM}}{m_X}\Omega_X h^2,
\label{OmegasuperWIMP}
\end{equation}
 where $\Omega_X h^2$ would be the thermal relic abundance 
 of the WIMP given by Eq.~(\ref{StandardThAbundance}) 
 if it were stable.

There are two typical superWIMP scenarios 
 in the context of supersymmetric models. 
One is a scenario realizing the axino as the LSP 
 with interactions suppressed by the Peccei-Quinn (PQ) 
 symmetry breaking scale~\cite{Covi:1999ty} . 
The other is a scenario where the gravitino is the LSP
 with Planck scale suppressed interactions~\cite{Feng:2003xh}.

The superWIMP scenario is an interesting possibility; 
 however, it is not so easy to consistently realize the scenario 
 with big bang nucleosynthesis (BBN). 
In supersymmetric models, 
 the next-to-LSP (NLSP) of, e.g., neutralino, stau, sneutrino  
 decays into the LSP and the standard model (SM) particles at late times. 
If the NLSP decays after BBN, its energetic daughters of the SM particles  
 would destroy light nuclei and spoil the success 
 of the standard BBN predictions. 
Furthermore, late-time injection of the energetic photon 
 from the NLSP decay would distort the spectrum of the observed 
 cosmic microwave background.

In the case of the axino LSP dark matter, 
 the constraints from BBN can be rather easily avoided 
 because the NLSP has a relatively 
 short lifetime~\cite{Covi:1999ty} and decays well before BBN.  
Some variants of this scenario are also possible~\cite{Nonthermal}.

In contrast to the axino LSP scenario, 
 for the case of gravitino LSP dark matter, 
 the constraints from BBN turn out to be very stringent. 
This is because the NLSP is normally long-lived 
 through the Planck mass suppressed interaction 
 between the NLSP and the gravitino LSP. 
When the NLSP is the neutralino or the stau (or the slepton in general),
 this scenario
 is very hard~\cite{onBBN,catalyzed,Kawasaki:2007xb, Kanzaki:2006hm}.
In the stau NLSP case, so-called catalyzed-BBN due to the bound state 
 with charged NLSP makes the problem worse~\cite{catalyzed,Kawasaki:2007xb}.
On the other hand, the sneutrino NLSP has rather mild constraints from BBN
 compared with the others.
Nevertheless, the sneutrino NLSP has not been appealing 
 from the beginning in this framework 
 because the thermal relic abundance $\Omega_{\tilde{\nu}}h^2$ is,
 as is well-known, too small to be suitable
 in Eq.~(\ref{OmegasuperWIMP}) due to the very efficient annihilation, 
 unless the NLSP sneutrino mass is of order of TeV~\cite{Falk:1994es}.
Such a heavy NLSP is not appealing 
 from the weak scale supersymmetry point of view. 
In order to obtain a suitable relic density of sneutrino NLSP 
 in Eq.~(\ref{OmegasuperWIMP}), 
 one may consider a nonthermal production of sneutrino NLSP 
 (see, for example, Ref.~\cite{Qball3/2}). 
However, in such a nonthermal scenario, 
 $\Omega_X h^2$ generally depends on many parameters in the scenario. 
Here, we stress that the sneutrino NLSP scenario is not suitable 
 as the superWIMP scenario, because of its too small thermal relic 
 abundance with a mass of ${\cal O}(100$ GeV), 
 rather than the BBN constraints.

In general, if we abandon providing the gravitino LSP relic density 
 from the NLSP decay through the formula in Eq.~(\ref{OmegasuperWIMP}), 
 we may be able to avoid the BBN constraints 
 in some parameter region in a model. 
In this case, the gravitino LSP should be produced 
 through scattering and decay processes of particles 
 in thermal plasma (thermal production). 
The relic density of the gravitino through thermal production 
 is evaluated as \cite{TP} 
\begin{eqnarray} 
\Omega^{\rm TP} h^2 \sim 0.27 
 \left( \frac{T_R}{10^{10}\; {\rm GeV}} \right ) 
 \left( \frac{100\; {\rm GeV}}{m_{3/2}} \right ) 
 \left( \frac{M_3}{1 \; {\rm TeV}} \right )^2 ,  
\label{TP}
\end{eqnarray} 
where $T_R$ is the reheating temperature after inflation, 
 and $M_3$ is the running gluino mass.  
By adjusting the reheating temperature, gravitino mass, etc., 
 we can obtain the relic abundance consistent with the WMAP data. 
However, this scenario may not be appealing 
 because the dark matter relic density has nothing to do 
 with the NLSP relic density.

In this paper, we consider the superWIMP scenario 
 with the gravitino LSP in the context of 
 nonstandard cosmological modes with 
 a modified expansion rate of the early Universe. 
The nonstandard cosmological effects cause dramatic changes 
 into the thermal relic density of the NLSP 
 with mass of ${\cal O}(100$ GeV), 
 so that this NLSP decay can yield the correct amount 
 of the gravitino dark matter relic density. 
We find that this scenario can be realized 
 consistently with the successful predictions of BBN. 
To be concrete, we consider the brane world cosmology~\cite{braneworld}  
 based on the picture proposed by Randall-Sundrum~\cite{RS}. 
Other cosmological models 
 such as quintessence scenarios~\cite{Salati:2002md}
 and scalar-tensor theories~\cite{Catena:2004ba} 
 have similar effects. 
In the brane world cosmology, 
 we examine two possibilities for the NLSP, 
 stau (or slepton, in general), and sneutrino.

\section{Brane world cosmology} 
Here we give a very brief review on our nonstandard 
 cosmological model, namely, the brane world cosmology, 
 which has been intensively investigated~\cite{braneworld}. 
This model is a cosmological version of the so-called RS II model 
 proposed by Randall and Sundrum~\cite{RS}. 
In the model, our four-dimensional universe is realized 
 on the 3-brane with a positive tension 
 located at the ultraviolet (UV) boundary 
 of a five-dimensional anti-de sitter spacetime. 
In this setup, the Friedmann equation for a spatially flat spacetime
 is given by 
\begin{equation}
H^2 = \frac{8\pi G}{3} \rho \left(1+\frac{\rho}{\rho_0}\right) ,
\label{BraneFriedmannEq}
\end{equation}
where
\begin{eqnarray}
\rho_0 = 96 \pi G M_5^6,
\end{eqnarray}
with $M_5$ being the five-dimensional Planck mass. 
Here the four-dimensional cosmological constant has been 
 tuned to be zero, and the negligible so-called dark radiation 
 has been omitted~\cite{rho^2}. 
The second term proportional to $\rho^2$
 is a new ingredient in the brane world cosmology
 and leads to a nonstandard expansion law of the early Universe. 
In the following, we call a temperature $T_t$ defined as 
 $\rho(T_t)=\frac{\pi^2}{30}g_* T_t^4 \equiv \rho_0$ ``transition temperature.''
This modification of the expansion law  $H^2 \propto \rho^2$
 at a high temperature ($T > T_t$) leads to some drastic changes
 for several cosmological issues, not only the thermal relic abundance 
 of dark matter~\cite{Okada:2004nc, Nihei:2004xv, 
 Nihei:2005qx,  OtherBraneDM} 
 but also other subjects, e.g., leptogenesis~\cite{Leptogenesis}. 
Needless to say, until the onset of BBN, 
 the standard expansion law, $H^2 \propto \rho$, 
 has to be recovered. 
This requirement leads to the model-independent constraint 
 on the transition temperature as $T_t \gtrsim 1$ MeV 
 or equivalently $M_5 \gtrsim 8.8$ TeV~\cite{braneworld}.

To be precise, we specify our setup here. 
We discuss a supersymmetric dark matter scenario, and 
 our model should be a supersymmetric version 
 of the Randall-Sundrum brane world model~\cite{RSsugra}. 
Since the Einstein equation belongs to the bosonic part 
 in supergravity, the cosmological solution 
 of Eq.~(\ref{BraneFriedmannEq}) remains the same. 
Although the gravitino is a bulk field, 
 we can, as a good approximation, regard it as 
 a field residing on the UV brane 
 as in the previous works~\cite{Okada:2004mh, gravitinoBW, Panotopoulos:2007fg}
 because the zero-mode gravitino is localized around 
 the UV brane as the same as the zero-mode graviton is. 
We assume the cancellation of the four-dimensional 
 cosmological constant by some mechanism different 
 from that discussed in the original paper~\cite{RS},  
 where the model parameter is severely constrained 
 by precision measurements of the gravitational law 
 in sub millimeter range, $T_t \gtrsim 1.3$ TeV 
 or equivalently $M_5 \gtrsim 1.1 \times 10^8$ GeV.  
This result is quite model dependent. 
For example, in dilatonic brane world models~\cite{maedawands}, 
 which seem to be a more natural five-dimensional effective theory 
 of an underlying fundamental theory than the original RS model, 
 the constraint can be very much moderated depending on the stabilization 
 mechanism of the dilaton field. 
Therefore, we take only the BBN bound into account 
 to keep our discussion general. 
Moreover, in this paper, we do not specify 
 the origin of supersymmetry breaking and its mediation to 
 the visible sector, 
 so that we treat the NLSP mass of ${\cal O}(100$ GeV) and the gravitino mass
 as a free parameter.

\section{Gravitino abundance from enhanced relic density}
As has been pointed out in Ref.~\cite{Okada:2004nc}, 
 the relic density of dark matter can be enhanced  
 in brane world cosmology 
 when the five-dimensional Planck mass $M_5$ is low enough.  
Applying this effect to the abundance of NLSPs, 
 we calculate the relic density of the gravitino LSP 
 produced by the decay of the NLSP at late times.

In the context of brane world cosmology, 
 the thermal relic density of a stable or a long-lived particle 
 by solving the Boltzmann equation,  
\begin{equation}
\frac{d n}{d t}+3Hn = -\langle\sigma v\rangle(n^2-n_{\rm{eq}}^2),
\label{n;Boltzmann}
\end{equation}
 with the modified Friedmann equation, Eq.~(\ref{BraneFriedmannEq}),
 where $n$ is the actual number density of the particle, 
 and $n_{\rm{eq}}$ is the equilibrium number density.  
If the particle freezes out at the brane world cosmology era 
 where $\rho^2$ dominates in Eq.~(\ref{BraneFriedmannEq}), 
 we can solve the Boltzmann equation analytically and find 
 an approximation formula~\cite{Okada:2004nc}, 
\begin{eqnarray}
 Y \equiv \frac{n}{s} 
 &\simeq& 0.54 \frac{x_t}{\lambda \langle \sigma v \rangle} . 
\label{Ybrane}
\end{eqnarray}
Here $\lambda = 0.26 (g_{*S}/g_*^{1/2}) M_{Pl} m$, 
 $m$ is the mass of the particle, $x_t$ is defined as $x_t \equiv m/T_t$, 
 and we have assumed the annihilation processes in S-wave. 
Note that the relic abundance is characterized 
 by the transition temperature rather than the decoupling temperature.

Using the well-known approximate formula 
 in the standard cosmology~\cite{Kolb}, 
\begin{eqnarray}
Y  &\simeq & \frac{x_d}{\lambda \langle \sigma v \rangle}   
\label{Ystandard}
\end{eqnarray}
with the freeze-out temperature $x_d = \frac{m}{T_f}$, 
 we obtain the ratio of the relic energy density of dark matter 
 in the brane world cosmology ($\Omega_{(b)}$) 
 to the one in the standard cosmology ($\Omega_{(s)}$) 
 such that
\begin{eqnarray}
\frac{\Omega_{(b)}}{\Omega_{(s)}}
 &\simeq& 0.54 \left(  \frac{x_t}{x_{d (s)}}  \right) , 
 \label{OmegaRatio} 
\end{eqnarray}
where $x_{d (s)}$ denotes the freeze-out temperature 
 in the standard cosmology. 
Therefore, the relic energy density in the brane world cosmology
 can be enhanced from the one in the standard cosmology  
 when the transition temperature is low enough, $T_t < T_f$. 
In the following analysis, we assume 
 such a low transition temperature.

\subsection{Slepton (stau) NLSP}
First, let us consider the case of the slepton NLSP. 
Among many possible channels of annihilation processes, 
 the dominant one is the annihilation into $\gamma\gamma$ and $Z\gamma$.
Hence, the cross section can be roughly given as ~\cite{AHS} 
\begin{equation}
 \langle\sigma v\rangle \simeq 
 \frac{4 \pi \alpha^2_{\rm em}}{m_{\tilde{l}}^2} .
\end{equation}
With this annihilation cross section, 
 the relic abundance of the slepton (if it were stable) 
 in the standard cosmology is expressed as~\cite{AHS} 
\begin{eqnarray} 
m_{\tilde{l}}Y_{\tilde{l}} \simeq 10^{-11} 
 \left(\frac{m_{\tilde{l}}}{100 \; {\rm GeV}}\right)^2 {\rm GeV} ,
\label{OmegaSlep2}
\end{eqnarray} 
 or in terms of the density parameter, 
\begin{eqnarray}  
\Omega_{\tilde{l}}h^2 
 \simeq \left(\frac{m_{\tilde{l}}}{2\; {\rm TeV}}\right)^2.   
\label{OmegaSlep}
\end{eqnarray}

Using Eqs.~(\ref{OmegaRatio}), (\ref{OmegaSlep2}) and (\ref{OmegaSlep}), 
 the slepton NLSP abundance in the brane world cosmology is modified into 
\begin{eqnarray}
m_{\tilde{l}}Y_{\tilde{l}(b)} &\simeq&  4\times 10^{-10} 
 [{\rm GeV}] \times 
 \left(\frac{m_{\tilde{l}}}{100\; {\rm GeV}}\right)^2 
 \left( \frac{23}{x_{d (s)}} \right)\left( \frac{x_t}{1700} \right), 
\label{Abundance} 
\\ 
{\Omega_{\tilde{l}(b)}} h^2 &\simeq&  0.1 \times
 \left( \frac{m_{\tilde{l}}}{100 \textrm{GeV}} \right)^2 
 \left( \frac{23}{x_{d (s)}} \right)
 \left( \frac{x_t}{1700}  \right)  .
\end{eqnarray}
Thus, the gravitino abundance from the slepton LSP decay 
 is given as 
\begin{eqnarray}
\Omega_{3/2}h^2 = \frac{m_{3/2}}{m_{\tilde{l}}}\Omega_{\tilde{l}(b)} h^2
 \simeq  0.1 \times \left(\frac{m_{3/2}}{m_{\tilde{l}}}\right)
 \left( \frac{m_{\tilde{l}}}{100 \; \textrm{GeV}} \right)^2 
 \left( \frac{23}{x_{d (s)}} \right)
 \left( \frac{x_t}{1700}  \right). 
\label{gravitinoAb}
\end{eqnarray} 
Although appropriate parameters enable us to obtain 
 the correct relic density of the dark matter gravitino, 
 these parameters are severely constrained by BBN. 
The consistency with the constraints from BBN is the key 
 whether the superWIMP scenario with the gravitino LSP 
 can be realized or not.

Now let us consider the constraints from BBN.  
Detailed analysis for the stau NLSP case has been done, 
 and the results are summarized in Figs.~2 and 3 
 for $m_{\tilde{\tau}_R}= 100$ GeV and $300$ GeV, respectively, 
 in Ref.~\cite{Kawasaki:2007xb} . 
In fact, we can see a large parameter region is excluded. 
For $m_{\tilde{\tau}_R}= 100$ GeV, 
 the region, which is consistent with BBN 
 and provides a suitable amount of the gravitino relic density 
 via the stau NLSP decay, can be read off as 
\begin{equation}
  (m_{\tilde{\tau}_R}Y_{\tilde{\tau}_R}, m_{3/2}) 
   \simeq (10^{-6}\, {\rm GeV}, \; 10 \; {\rm MeV}). 
\label{100} 
\end{equation}
In order to realize this allowed region, 
 Eq.~(\ref{Abundance}) indicates $ T_t \simeq 24$ keV 
 for $x_{d(s)} \sim 23$, so that this result contradicts against 
 the requirement $T_t \gtrsim 1$ MeV. 
For $m_{\tilde{\tau}_R}= 300$ GeV, the allowed region appears in  
\begin{eqnarray}
 m_{\tilde{\tau}_R}Y_{\tilde{\tau}_R} 
 \gtrsim 2 \times 10^{-7} \, {\rm GeV}, 
 \; \;  m_{3/2} \lesssim 0.6 \,{\rm GeV}. 
\label{300}
\end{eqnarray}
We find that 
 the transition temperature, $T_t \lesssim 3.2$ MeV 
 (or equivalently $M_5 \lesssim 19$ TeV),  
 can realize this region. 
This result is close to the BBN bound on the transition temperature 
 $T_t \gtrsim 1$ MeV, 
 and the gravitino superWIMP scenario with the stau NLSP 
 is marginally possible in the brane world scenario. 
Note that the standard cosmology cannot realize this scenario 
 as can be seen from Eq.~(\ref{OmegaSlep}) 
 and the enhancement of the stau NLSP abundance 
 by the brane world cosmological effects is necessary.

In the allowed region we have considered, 
 the gravitino is as light as sub-GeV 
 while the NLSP is as heavy as $100$ GeV. 
Hence, this gravitino dark matter produced by the decay of stau 
 may have a large velocity dispersion. 
While cold dark matter (CDM) has made great success 
 in the structure formation of the Universe at a large scale ($\gtrsim 1$ Mpc), 
 the recent high-resolution $N$-body simulations 
 of the CDM showed some discrepancies at a small scale ($\lesssim 1$ Mpc). 
One is the so-called ``missing satellite 
 problem''~\cite{MissingSatellite}; namely, 
 the $N$-body simulation predicts more virialized dark objects 
 with much smaller mass scale than observed. 
The other is that CDM models predict a cuspier mass distribution 
 of CDM halos than observed: ``cusp problem''~\cite{CuspProblem}. 
It has been discussed~\cite{warmsuperWIMP, HisanoInoueTakahashi} 
 that these problems can be solved in the superWIMP scenario 
 with appropriate large velocity dispersion of the superWIMP, 
 which reduces the number of less massive subhalos and  
 the primordial fluctuations at the small scales;
 its difficulty is also pointed out though~\cite{Bringmann:2007ft}. 
On the other hand, too large velocity dispersion of 
 the superWIMP dark matter, 
 which exceeds the damping scale $\gtrsim 1$ Mpc, 
 is constrained from Lyman alpha clouds. 
The damping scale can be calculated 
 for a given lifetime of NLSP~\cite{Feng:2003uy} 
\begin{equation}
\tau_{\rm NLSP} \simeq 48\pi M_P^2 \frac{m_{3/2}^2}{m_{\rm NLSP}^3} ,
\end{equation}
 with $M_P$ being the reduced Planck mass, 
 the NLSP mass $m_{\rm NLSP}$ and the gravitino mass $m_{3/2}$, 
 and a given mass ratio $m_{3/2}/m_{\rm NLSP}$ 
 (see Fig.~1 in Ref.~\cite{HisanoInoueTakahashi}).  
In our case with $m_{3/2}$ in sub-GeV and slepton NLSP mass 
 around 100 GeV, typical damping scale is found to be 
 ${\cal O}(1\; {\rm Mpc})$.  
Hence, our scenario is in fact marginally possible. 

In our analysis, we have only considered the gravitino abundance 
 from the decay of the NLSP. 
In general, the gravitino production from thermal plasma 
 should be considered. 
As has been pointed out in Ref.~\cite{Okada:2004mh}, 
 the relic density of the gravitino through the thermal production 
 in the brane world cosmology can be obtained 
 from Eq.~(\ref{TP}) with the replacement $T_R \rightarrow 2 T_t$ 
 in the case with $T_t \ll T_R$, 
\begin{eqnarray} 
\Omega^{\rm TP}_{(b)} h^2 &\sim&  0.54 
 \left( \frac{T_t}{10^{10}\; {\rm GeV}} \right ) 
 \left( \frac{100\; {\rm GeV}}{m_{3/2}} \right ) 
 \left( \frac{M_3}{1 \; {\rm TeV}} \right )^2 \nonumber \\
&=& 0.54 \times 10^{-8} 
 \left( \frac{1}{x_t} \right) 
 \left( \frac{m_{\tilde{l}}}{m_{3/2}} \right ) 
 \left( \frac{M_3}{1 \; {\rm TeV}} \right )^2.    
\label{TPB}
\end{eqnarray} 
The gravitino density from the thermal production 
 is negligible unless $T_R \gtrsim 10^8$ GeV.  
This is also an important effect in the brane world cosmology. 
The gravitino abundance given in Eq.~(\ref{gravitinoAb}) 
 is reduced into the standard cosmological one for 
 $x_t \leq x_{d(s)}$, 
\begin{eqnarray} 
\Omega_{3/2}h^2 
 \simeq  0.25 \times 10^{-2} 
 \left(\frac{m_{3/2}}{m_{\tilde{l}}}\right)
 \left( \frac{m_{\tilde{l}}}{100 \; \textrm{GeV}} \right)^2,  
\label{gravitinoSTD}
\end{eqnarray} 
where we have used Eq.~(\ref{OmegaSlep}). 
For $m_{\tilde{l}} \sim 100$ GeV in our interest, 
 the gravitino relic abundance in the standard cosmology 
 is too small to be consistent with the WMAP data as mentioned above. 
As the transition temperature is raised 
 from $T_t \sim m_{\tilde{l}}/1700$ [see Eq.~(\ref{gravitinoAb})], 
 the total gravitino relic density from the sum of
 the NLSP decay and the thermal production is decreasing
 and reaches the standard cosmological value. 
When the transition temperature is raised further, 
 the gravitino relic density from the thermal production 
 exceeds that from the NLSP decay for $T_t \gtrsim 10^6$ GeV 
 and becomes consistent with the WMAP data at $T_t \sim 10^8$ GeV.

\subsection{Sneutrino NLSP}

In the standard cosmology, the density parameter for 
 the relic sneutrino is estimated as~\cite{Falk:1994es}
\begin{equation}
\Omega_{\tilde{\nu}}h^2 \simeq 0.2 \times
 \left(\frac{m_{\tilde{\nu}}}{1 \; \textrm{TeV}}\right)^2.
\label{OmegaSnu}
\end{equation}
Together with Eqs.~(\ref{OmegaRatio}) and (\ref{OmegaSnu}), 
 we find, in the brane world cosmology, 
\begin{equation}
 {\Omega_{\tilde{\nu}(b)}} h^2 \simeq 0.1 \times
 \left( \frac{m_{\tilde{\nu}}}{100 \; \textrm{GeV}} \right)^2 
 \left( \frac{23}{x_{d (s)}} \right)
 \left( \frac{x_t}{2100}  \right)  ,
\end{equation}
or equivalently
\begin{eqnarray}
 m_{\tilde{\nu}}Y_{\tilde{\nu}(b)} \simeq 
 4\times 10^{-10} \; [{\rm GeV}]  \times  
 \left(\frac{m_{\tilde{\nu}}}{100 \; {\rm GeV}}\right)^2 
 \left( \frac{23}{x_{d (s)}} \right)\left( \frac{x_t}{2100} \right).  
\label{AbunSnu}
\end{eqnarray}
Therefore, due to the enhancement by the modified expansion law, 
 the sneutrino with mass ${\cal O}$(100 GeV) 
 can provide a suitable relic density 
 of the gravitino LSP via its decay. 
The resultant density parameter of gravitino is given as
\begin{eqnarray}
\Omega_{3/2}h^2 = \frac{m_{3/2}}{m_{\tilde{\nu}}}\Omega_{\tilde{\nu}} h^2
 \simeq  0.1 \times \left(\frac{m_{3/2}}{m_{\tilde{\nu}}}\right)
 \left( \frac{m_{\tilde{\nu}}}{100 \; \textrm{GeV}} \right)^2 
 \left( \frac{23}{x_{d (s)}} \right)
 \left( \frac{x_t}{2100}  \right)
\end{eqnarray}
 from Eq.~(\ref{OmegasuperWIMP}).

The BBN constraints for the sneutrino LSP 
 have been analyzed in detail, 
 and the results are summarized in Fig.~5 
 for $m_{\tilde{\nu}} =300$ GeV in Ref.~\cite{Kanzaki:2006hm} .  
The allowed region can be read off as 
\begin{eqnarray}
{\rm (i)} & m_{\tilde{\nu}}Y_{\tilde{\nu}(b)} 
 \gtrsim 2\times 10^{-6}\, {\rm GeV}, 
  \; &  m_{3/2} \lesssim 0.06 \, {\rm GeV} , 
\nonumber    \\
{\rm (ii)} & m_{\tilde{\nu}}Y_{\tilde{\nu}(b)} 
 \simeq 2 \times 10^{-7}\, {\rm GeV}, 
  \; &  m_{3/2} \simeq 0.3 \, {\rm GeV} , 
\nonumber  \\
{\rm (iii)}& 
   4 \times 10^{-10}\, {\rm GeV} \lesssim 
 m_{\tilde{\nu}}Y_{\tilde{\nu}(b)} \lesssim 
   7 \times 10^{-10}\, {\rm GeV}, 
  \; & 300 \; {\rm GeV} >  m_{3/2} \gtrsim 150\, {\rm GeV} . 
\nonumber 
\end{eqnarray}
For each allowed region, Eq.~(\ref{AbunSnu}) leads to 
\begin{eqnarray}
{\rm (i)} &T_t \lesssim 0.26 \; {\rm MeV} &
\; \; ( M_5 \lesssim 3.6 \; {\rm TeV}), \nonumber \\
{\rm (ii)} & T_t \simeq  2.6 \; {\rm MeV} &
\; \; ( M_5 \simeq 17 \; {\rm TeV}), \nonumber \\
{\rm (iii)} & 1.3 \; {\rm GeV} >   T_t \gtrsim 730 \; {\rm MeV} &
\; \; ( 1000 \; {\rm TeV} > M_5 \gtrsim 710 \; {\rm TeV}). \nonumber 
\end{eqnarray} 
Region (i) is not consistent with $T_t \gtrsim 1$ MeV 
 and hence excluded. 
Point (ii) is marginally acceptable 
 as in the case of the stau NLSP but similarly delicate
 from viewpoints of both BBN and free-streaming constraints. 
Finally, region (iii) is typical for the sneutrino NLSP case, 
 because this appears when masses of the sneutrino NLSP 
 and the gravitino LSP are close 
 and such a region is never allowed for other NLSP cases. 
In addition, one may notice that the free-streaming length of 
 region (iii) is less than a Mpc scale unlike point (ii).

\section{Conclusion and discussions}
We have investigated that gravitino dark matter 
 produced by the decay of thermal relic NLSP, 
 in particular, stau NLSP and sneutrino NLSP, 
 in the context of brane world cosmology 
 \footnote{
 Recently, the gravitino dark matter in brane world cosmology 
 have been investigated in Ref.~\cite{Panotopoulos:2007fg}. 
 In the paper, the NLSP is identified as the neutralino in the framework of 
 the constrained minimal supersymmetric standard model (CMSSM) 
 and the thermal production of gravitino is also taken into account. 
}. 
Even if the NLSP mass is around 100 GeV, 
 the brane world cosmological effects can dramatically enhance  
 thermal relic density of the NLSP enough to yield the correct 
 abundance for the gravitino dark matter through the NLSP decay. 
We have found that with a suitable five-dimensional Planck mass, 
 this scenario can be consistent with BBN constraints. 
The brane world cosmological effects 
 suppress the thermal production of the gravitino, 
 so that the energy density from the thermal production 
 is negligible in our scenario.

In the brane world cosmology, a disfavored candidate of dark matter 
 due to too efficient annihilation in the standard cosmology
 can be a suitable one 
 with the help of the enhancement effect for its relic density. 
The case of the winolike neutralino dark matter 
 has been analyzed in detail in Ref.~\cite{Nihei:2005qx}, 
 and it has been found that the winolike neutralino 
 with mass around 100 GeV can be consistent with the WMAP data. 
We have shown that the same enhancement mechanism 
 is applicable to the NLSP 
 in the superWIMP scenario with the gravitino LSP. 
The sneutrino has no electric charge and is a potential candidate 
 of the dark matter, if its relic density is enhanced 
 as in the case of winolike dark matter in Ref.~\cite{Nihei:2005qx}. 
However, the possibility of sneutrino dark matter has been excluded 
 by the null result in direct dark matter searches.  
The sneutrino NLSP in the superWIMP scenario 
 would be interesting for the following reason: 
There is no relic sneutrino in the present Universe 
 and the sneutrino cannot be observed in direct dark matter searches, 
 nevertheless the gravitino dark matter inherited 
 its relic density from that of the sneutrino. 
In addition, the sneutrino is a harmless mother particle 
 decaying into gravitino dark matter 
 rather than others like the neutralino or slepton, 
 because of its weak constraints from BBN.

When the transition temperature or equivalently $M_5$ is very low, 
 the enhancement of thermal relic density of a particle 
 would be much too to be consistent with the WMAP data. 
This fact implies a lower bound on $T_t$ or $M_5$, 
 which can be stronger than the model-independent BBN bound, 
 $T_t \gtrsim 1$ MeV or equivalently $M_5 \gtrsim 8.8$ TeV.  
In fact, the lower bound on $M_5 \gtrsim 600$ TeV 
 has been reported for neutralino dark matter 
 in the CMSSM~\cite{Nihei:2004xv}, 
 while $M_5 \gtrsim 100$ TeV 
 for the winolike dark matter~\cite{Nihei:2005qx}.  
In the superWIMP scenario we have discussed, 
 the factor, $m_{3/2}/m_{NLSP}$,  in Eq.~(\ref{OmegasuperWIMP}) 
 can play an important role so as to moderate too much enhanced 
 NLSP energy density and as a result, the lower bound 
 on $T_t$  or $M_5$ can be as low as the model-independent BBN bound. 
Of course, if the factor $m_{3/2}/m_{NLSP}$ is too small,
 the resultant dark matter becomes too warm as we have discussed.

The long-lived NLSP is also interesting at future collider experiments 
 such as the Large Hadron Collider (LHC) and 
 the International Linear Collider (ILC). 
In the superWIMP scenario with the gravitino LSP, 
 the decay length of the NLSP well exceeds the detector size 
 of the LHC and the ILC, and the NLSP decay takes place 
 outside the detector. 
In this case, there have been interesting proposals~\cite{sWIMPcollider}
 for the way to trap quasistable NLSPs outside the detector, 
 when the NLSP is a charged particle like stau. 
Detailed studies of the NLSP decay may provide 
 precise measurements of the gravitino mass 
 and the four-dimensional Planck mass. 
In our brane world scenario, 
 such precise measurements can also give the information of $M_5$.

If the long-lived NLSP is a charge neutral particle such as the neutralino 
 and the sneutrino, it is very hard to observe its decay. 
However, in this case, it is important to distinguish the sneutrino from the neutralino. 
At collider experiments, the long-lived sneutrino behaves 
 similarly to the LSP (or long-lived NLSP) neutralino, 
 but there are some characteristic signatures~\cite{Covi:2007xj}. 
Interestingly, once the sneutrino has been observed 
 as if it were the LSP at collider experiments, 
 the superWIMP scenario would become more plausible 
 as the dark matter scenario. 
This is because a simple scenario with the sneutrino LSP 
 has already been excluded by direct dark matter search experiments 
 and the sneutrino must be unstable, 
 if the LSP is the dominant component of dark matter. 
As in the case of stau NLSP, 
 precise measurements of sneutrino properties 
 can be interpreted into the information of $M_5$.

If the neutralino has been observed as the LSP at collider 
 experiments and its properties have been precisely measured, 
 one can calculate the neutralino relic density 
 at the present Universe.  
Suppose that the result shows too large relic density 
 inconsistent with the WMAP data. 
In this case, the superWIMP scenario can provide a solution 
 to this contradiction: 
The neutralino is the NLSP and decays into the gravitino LSP, 
 whose energy density is reduced by the factor, 
 $m_{3/2}/m_{NLSP}$, in Eq.~(\ref{OmegasuperWIMP}). 
However, the key to realize this scenario 
 lies on how to avoid the constraints from BBN.

%
\section*{Acknowledgments}
The work of N.O. is supported in part 
 by the Grant-in-Aid for Scientific Research 
 from the Ministry of Education, Science and Culture of Japan, 
 (No.~18740170). 
The work of O.S. is supported by the MEC Project FPA 2004-02015, 
 the Comunidad de Madrid Project HEPHACOS (No.~P-ESP-00346),
 and the European Network of Theoretical Astroparticle Physics ILIAS/ENTApP 
 under contract number RII3-CT-2004-506222. 



\end{document}